\documentstyle[11pt]{article}
\newcommand{\be}{\begin{equation}}
\newcommand{\en}{\end{equation}}
\newcommand{\bea}{\begin{eqnarray}}
\newcommand{\ena}{\end{eqnarray}}

\newcommand{\vs}[1]{\rule[ - #1 mm]{0 mm}{#1 mm}}
\newcommand{\Q}{{\mbox{$Q$}}}
\newcommand{\W}{{\sf W}}
\newcommand{\NP}[1]{Nucl.\ Phys.\ {\bf #1}}
\newcommand{\PL}[1]{Phys.\ Lett.\ {\bf #1}}

\newcommand{\CMP}[1]{Comm.\ Math.\ Phys.\ {\bf #1}}

\newcommand{\MPL}[1]{Mod.\ Phys.\ Lett.\ {\bf #1}}
\newcommand{\IJMP}[1]{Int.\ J.\ Mod.\ Phys.\ {\bf #1}}

\begin{document}
%
%
\renewcommand{\thefootnote}{\fnsymbol{footnote}}
\newpage
\setcounter{page}{0}
\pagestyle{empty}
%
%
\rightline{DFTT-25/93}
\rightline{hep-th/9306019}
\rightline{May 1993}

\vs{15}

\begin{center}
{
\LARGE {Explicit Construction of the BRST Charge for \W$_4$  }
 }\\[1cm]
{\large K.\ Hornfeck}\footnote{e-mail: HORNFECK@TO.INFN.IT; \hspace{0.5cm}
31890::HORNFECK}\\[0.5cm]
{\em INFN, Sezione di
Torino, Via Pietro Giuria 1, I-10125 Torino,
Italy}
\\[1cm]
\end{center}
\vs{15}

\centerline{ \bf{Abstract}}
We give the explicite form of the BRST charge $\Q$ for the algebra  \W$_4 =
$ {\sf WA}$_3$ in the basis where the spin-3 and the spin-4 field are
primary as well as for a basis where the algebra closes quadratically.
%
%
\renewcommand{\thefootnote}{\arabic{footnote}}
\setcounter{footnote}{0}
\newpage
\pagestyle{plain}
For the construction of {\sf W}-strings~\cite{BG89}, especially its
physical states, it is important to know  the  cohomology of the BRST
charge~\cite{DDR91,Ram91,PRS91,PSSW92,LNP92,Wes92,LPSW93,FW93}.
Unfortunately,  for the {\sf W}-strings based on the algebras \W$_n$ = {\sf
WA}$_{n-1}$ only the one for the simplest algebra, the Zamolodchikokv
\linebreak
{\sf W}$_3$-algebra~\cite{Zam85} has been  constructed
previously~\cite{ThM87}. Since so far there is no simplifying  ansatz of
the  terms appearing in \Q for a given (general) \W-algebra~\footnote{as it
exists for a  certain sub-class of \W-algebras that close
quadratically~\cite{SSvN89}.}, we  constructed the BRST  charge for the
\W$_4$-algebra by taking into  account all possible terms of even parity
(where the matter field $W_3$ and the ghost fields $c_3$ and $b_3$ have
parity odd, all other fields have parity even) and determining the
coefficients by demanding that \Q\ is nilpotent. However, contrary to the
case of the \W$_3$-algebra, one has to be careful about two points:

We have the freedom of redefining the spin-4 field $W_4$ (that is usually
taken to be quasi-primary) by transforming $W_4 \rightarrow  W_4 + \kappa\,
\Lambda$, $\Lambda$ being the spin-4 quasi-primary $T T - 3/10\, T''$ (all
products of fields are considered to be normal ordered in the standard way
whenever necessary).  There are two obvious choices: Either $W_4$ is such
that the algebra closes quadratically or else $W_4$ is taken to be primary,
leading also to cubic terms (in the Virasoro operator $T$) in the algebra.
Whereas in the latter  case  the operator product expansions (OPEs) (and
therefore the commutators) are unique (apart from a sign-ambiguity due to
the transformation $W_4 \rightarrow - W_4$, there exist two (equivalent)
possibilities for the algebra that closes  quadratically. The structure
constants of~\W$_4$  can be found  in refs.~\cite{BFK91,KW91}.

To the matter part ($T, W_3, W_4$) we have to introduce ghost and
anti-ghost fields $(\{c_2, b_2\}, \{c_3, b_3\},$ $\{c_4, b_4\})$, obeying
the OPEs
\bea
c_i \star c_j & = & b_i \star b_j \, = \, 0 \nonumber \\
c_i \star b_j & = & \delta_{i j} \,\,1
\label{ghostope}
\ena
and the BRST charge will be of the form
\bea
\Q & = & \oint dw \, \left(T \,c_2 + W_3 \,c_3 + (W_4 + \kappa_1 \,TT +
\kappa_2 \,
T'')\, c_4 \right) (w) \,\,
+ \\
&&\hspace{5mm} \mbox{contributions containing anti-ghost fields $b_i$}
\nonumber
\ena
We introduced $\kappa_1$ and $\kappa_2$ to include different bases for the
field $W_4$.

However, the OPEs (\ref{ghostope}) do not fix the ghosts in a unique way.
Indeed one might  think of changing the ghost and anti-ghost fields by a
``canonical transformation'' to
\bea
c_i \,\rightarrow \,\tilde{c}_i & = & c_i + f_i(\{c,b\}) \nonumber \\
b_i \, \rightarrow \, \tilde{b}_i & = & b_i + g_i(\{c,b\})
\label{ghosttrans}
\ena
where the functions $f_i$ and $g_i$ obbey the following rules: \newline
1) they have ghost-number $+1$ or $-1$, respectively;\newline
2) they have the right conformal dimension;\newline
3) they have the correct parity;\newline
4) the new ghosts $\tilde{c}$ and $\tilde{b}$ obbey exactly the same OPEs
   as (\ref{ghostope}).

These transformations are different from the change of basis
of~\cite{LPSW93}, since they involve only the ghost sector and do not mix
ghost and matter fields. In the \W$_3$-algebra one realizes very soon, that
there no such transformation exists. For \W$_4$, however, we can indeed
perform such a change of basis by a 8-parameter transformation. The
simplest one being
\be
\tilde{c}_2 = c_2 + \alpha \,c_4''; \hspace{2cm} \tilde{b}_4 = b_4 -
\alpha \,
b_2''
\label{simtrans}
\en
and leaving all other ghosts unchanged.

This special transformation would lead in \Q\ to a term of the kind $c_4 \,
T''$  and can  therefore be absorbed into $\kappa_2$, leaving seven free
parameters (apart from $\kappa_1$ and $\kappa_2$) that should show up
in~\Q, after having imposed that \Q$^2 = 0$. This is indeed the case.

The Virasoro operator $T_{\mbox{\small gh}}$ for the ghost-sector is
given in the form
\be
T_{\mbox{\small gh}} = (c_2 \,b_2' + 2 c_2' \,b_2) + (2 c_3 \, b_3' + 3
c_3' \, b_3) +
(3 c_4 \, b_4' + 4 c_4' \, b_4)
\label{ghostVir}
\en
with a central charge $c_{\mbox{\small gh}} = - 246$ such that the total
Virasoro  generator $T_{\mbox{\small total}} = T + T_{\mbox{\small  gh}}$
is anomaly free if the matter  Virasoro operator $T$ has central charge
$c_{\mbox{\small matter}} = 246$. It is clear that in general the
transformations (\ref{ghosttrans}) will  change the Virasoro operator
(\ref{ghostVir}) to $\widetilde{T}_{\mbox{\small gh}}$,  even  though
$\widetilde{T}_{\mbox{\small gh}}$ is again a Virasoro operator with the
correct central charge $-246$.

There is a four-dimensional subset, however, that leaves $T_{\mbox{\small
gh}}$ invariant under (\ref{ghosttrans}) and once  demanding that the
commutator of \Q\ with $b_2$ reproduces $T_{\mbox{\small total}}$ and
absorbing again one parameter into $\kappa_2$ we are finally left with two
open parameters.

For one of them there is a convenient way to fix by demanding that setting
the fields $W_4$, $c_4$ and $b_4$ formally to zero would reproduce the BRST
operator of the \W$_3$-algebra. Since we did not find such a convenient
choice for the second free parameter, we leave it open in the result,
denoted as $\beta$. This parameter is connected to the transformation
\bea
\tilde{c}_4 & = & { c_4}
\nonumber \\
\tilde{b}_4 & = & { b_4} - 4\,\beta \,\,{ }{ b_2}'\,
    { }{ b_2}\,{ }{ c_3}\,{ }{ b_3}\,{ c_4}' -
  2\,\beta \,\,{ }{ b_2}'\,
    { }{ b_2}\,{ }{ c_3}\,{ }{ b_3}'\,{ c_4} -
  2\,\beta \,\,{ }{ b_2}'\,
    { }{ b_2}\,{ }{ c_3}'\,{ }{ b_3}\,{ c_4} - \nonumber \\
&&  2\,\beta \,\,{ }{ b_2}''\,
    { }{ b_2}\,{ }{ c_3}\,{ }{ b_3}\,{ c_4} -
  2\,\beta \,\,{ }{ b_2}''\,{ }{ b_2}\,{ c_4}' -
  \beta \,\,{ }{ b_2}''\,{ }{ b_2}'\,{ c_4} -
  \beta \,\,{ }{ b_2}^{(3)}\,{ }{ b_2}\,{ c_4}
\nonumber \\
\tilde{c}_3 & = & { c_3} - 2\,\beta \,\,{ }{ b_2}'\,
    { }{ b_2}\,{ }{ c_3}\,{ }{ c_4}'\,{ c_4}
\nonumber \\
\tilde{b}_3 & = & { b_3} + 2\,\beta \,\,{ }{ b_2}'\,
    { }{ b_2}\,{ }{ b_3}\,{ }{ c_4}'\,{ c_4}
\\
\tilde{c}_2 & = & { c_2} - 2\,\beta \,\,{ }{ b_2}\,
    { }{ c_3}\,{ }{ b_3}\,{ }{ c_4}''\,{ c_4} -
  2\,\beta \,\,{ }{ b_2}\,{ }{ c_3}\,
     { }{ b_3}'\,{ }{ c_4}'\,{ c_4} -
  2\,\beta \,\,{ }{ b_2}\,{ }{ c_3}'\,
     { }{ b_3}\,{ }{ c_4}'\,{ c_4} + \nonumber \\
&&  \beta \,\,{ }{ b_2}\,{ }{ c_4}''\,{ c_4}' +
  \beta \,{ }{ b_2}\,{ }{ c_4}^{(3)}\,{ c_4} -
  4\,\beta \,\,{ }{ b_2}'\,
    { }{ c_3}\,{ }{ b_3}\,{ }{ c_4}'\,{ c_4} +
  2\,\beta \,\,{ }{ b_2}'\,{ }{ c_4}''\,{ c_4}
\nonumber \\
\tilde{b}_2 & = & { b_2}
\nonumber
\ena

Of course, one could also set all seven parameters to different values
(working with a different~$\widetilde{T}_{\mbox{\small  gh}}$ if
necessary), but the final result written in all open parameters is too
complicated to be presented here.

In the table are given the various fields $j_n$ contributing to the BRST
charge $\Q = \oint dw \sum j_n(w) \, a_n$ together with the coefficient
$a_n$ both for the primary basis as well as for one of the bases in which
the algebra closes quadratically and for that the structure constant
$(C_{44}^4)^2$ is given by $\left[ 54/5 \,(c+47)^2 /((c+22) \,(33+14\,c))
\right]$. In both cases we took the algebra with the positive root for
$C_{44}^4$.

We did not try to find a transformation similar to the one proposed
in~\cite{LPSW93} for \W$_3$, that splits~\Q\ into various anticommuting and
nilpotent parts of different ($c_2,c_3,c_4$)-ghost-number, but on the basis
of the result presented here it should not be too difficult to obtain.

The calculations have been performed using the {\sc Mathematica} package
for computing OPEs by K.~Thielemans~\cite{Thi91}. I am grateful to
M.~Freeman for discussions. \vs{15}

\large{Table of contributions to the BRST charge}

\begin{center}
\begin{tabular}{l|l|l}
$j_n$ & $a_n$, primary basis & $a_n$, quadratic basis \\ \hline
$ { }T\,{ c_2} $& $ 1
  $ & $ 1
  $ \\[3mm]
$ { }{ W_3}\,{ c_3}$& $ 1
  $ & $ 1
  $ \\[3mm]
$ { }{ W_4}\,{ c_4}$& $ 1
  $ & $ 1
  $ \\[3mm]
$ { }{ c_2}'\,{ }{ c_2}\,{ b_2}$& $ -1
  $ & $ -1
  $ \\[3mm]
$ { }{ b_2}\,{ }{ c_3}''\,{ c_3}'$& $ {{49}\over {626}}
  $ & $ -{{11706}\over {16445}}
  $ \\[1mm]
$ { }{ b_2}\,{ }{ c_3}^{(3)}\,{ c_3}$& $ -{{49}\over {939}}
  $ & $ {{7804}\over {16445}}
  $ \\[3mm]
$ { }T\,{ }{ b_2}\,{ }{ c_3}'\,{ c_3}$& $ -{4\over {313}}
  $ & $ -{5\over {253}}
  $ \\[3mm]
$ { }{ W_3}\,{ }{ b_2}'\,{ }{ c_3}\,{ c_4}$& $
-9\,{\sqrt{{{17}\over {19638872}}}}
  $ & $ {{-251\,{\sqrt{{5\over {155306}}}}}\over {13}}
  $ \\[1mm]
$ { }{ W_3}\,{ }{ b_2}\,{ }{ c_3}'\,{ c_4}$& $
{{-353}\over {{\sqrt{83465206}}}}
  $ & $ {{-5521}\over {13\,{\sqrt{776530}}}}
  $ \\[1mm]
$ { }{ W_3}\,{ }{ b_2}\,{ }{ c_3}\,{ c_4}'$& $
{{553}\over {3\,{\sqrt{83465206}}}}
  $ & $ {{711\,{\sqrt{{8\over {388265}}}}}\over {13}}
  $ \\[3mm]
$ { }{ b_2}\,{ }{ c_4}^{(5)}\,{ c_4}$& $ {{14\,\beta }\over
{15}}
  $ & $ {{14\,\beta }\over {15}}
  $ \\[1mm]
$ { }{ b_2}\,{ }{ c_4}^{(4)}\,{ c_4}'$& $ {{13\,\left(
15861457 + 5320659456\,\beta  \right) }\over
    {63847913472}}
   $ & $ {{-39373278251 + 152860862336\,\beta }\over {141102334464}}
  $ \\[1mm]
$ { }{ b_2}\,{ }{ c_4}^{(3)}\,{ c_4}''$& $ {{-\left( 818537
+ 591184384\,\beta  \right) }\over {506729472}}
  $ & $ {{6488840259 - 58792639360\,\beta }\over {50393690880}}
  $ \\[3mm]
$ { }T\,{ }{ b_2}''\,{ }{ c_4}'\,{ c_4}$& $ -{{19577}\over
{31670592}}
  $ & $ {{90912053}\over {1049868560}}
  $ \\[1mm]
$ { }T\,{ }{ b_2}\,{ }{ c_4}''\,{ c_4}'$& $ {{4055111 -
5320659456\,\beta }\over {5320659456}}
  $ & $ {{-\left( 5204269653 + 58792639360\,\beta  \right) }\over
    {58792639360}}
  $ \\[1mm]
$ { }T\,{ }{ b_2}\,{ }{ c_4}^{(3)}\,{ c_4}$& $ {{-\left(
7350521 + 5320659456\,\beta  \right) }\over {5320659456}}
  $ & $ {{18757441753 - 176377918080\,\beta }\over {176377918080}}
  $ \\[1mm]
$ { }T\,{ }{ b_2}'\,{ }{ c_4}''\,{ c_4}$& $ {{-\left(
1709695 + 1773553152\,\beta  \right) }\over {886776576}}
  $ & $ {{6316984623 - 58792639360\,\beta }\over {29396319680}}
  $ \\[3mm]
\end{tabular}
\end{center}
\newpage
\begin{center}
\begin{tabular}{l|l|l}
$ { }{ W_4}\,{ }{ b_2}\,{ }{ c_4}'\,{ c_4}$& $ 9\,
{\sqrt{{{17}\over {4909718}}}}
  $ & $ {{251\,{\sqrt{{{10}\over {77653}}}}}\over {13}}
  $ \\[3mm]
$ { }T\,{ }T\,{ }{ b_2}\,
     { }{ c_4}'\,{ c_4}$& $ -{{553}\over {2969118}}
  $ & $ 0
  $ \\[3mm]
$ { }{ c_2}'\,{ }{ c_3}\,{ b_3}$& $ -2
  $ & $ -2
  $ \\[1mm]
$ { }{ c_2}\,{ }{ c_3}'\,{ b_3}$& $ 1
  $ & $ 1
  $ \\[3mm]
$ { }{ c_3}^{(3)}\,{ }{ b_3}\,{ c_4}$& $ {{3297}\over
{{\sqrt{5341773184}}}}
  $ & $ {{-19693}\over {13\,{\sqrt{12424480}}}}
  $ \\[1mm]
$ { }{ c_3}''\,{ }{ b_3}\,{ c_4}'$& $ {{-13\,
{\sqrt{{{2048}\over {41732603}}}}}\over 3}
  $ & $ {{16979}\over {13\,{\sqrt{3106120}}}}
  $ \\[1mm]
$ { }{ c_3}'\,{ }{ b_3}\,{ c_4}''$& $ -91\,{\sqrt{{{17}\over
{314221952}}}}
  $ & $ {{-67533}\over {13\,{\sqrt{12424480}}}}
  $ \\[1mm]
$ { }{ c_3}\,{ }{ b_3}\,{ c_4}^{(3)}$& $ {{-7\,
{\sqrt{{{391}\over {13661824}}}}}\over 3}
  $ & $ {{-3949}\over {13\,{\sqrt{12424480}}}}
  $ \\[3mm]
$ { }T\,{ }{ c_3}\,{ }{ b_3}'\,{ c_4}$& $ 0
  $ & $ {{79\,{\sqrt{{{40}\over {77653}}}}}\over {13}}
  $ \\[1mm]
$ { }T\,{ }{ c_3}'\,{ }{ b_3}\,{ c_4}$& $ 0
  $ & $ {{-2133\,{\sqrt{{2\over {388265}}}}}\over {13}}
  $ \\[1mm]
$ { }T\,{ }{ c_3}\,{ }{ b_3}\,{ c_4}'$& $ 0
  $ & $ {{553\,{\sqrt{{{32}\over {388265}}}}}\over {13}}
  $ \\[3mm]
$ { }{ W_3}\,{ }{ b_3}\,{ }{ c_4}'\,{ c_4}$& $ -{{1565}\over
{50592}}
  $ & $ -{{18975}\over {621224}}
  $ \\[3mm]
$ { }{ c_2}'\,{ }{ c_4}\,{ b_4}$& $ -3
  $ & $ -3
  $ \\[1mm]
$ { }{ c_2}\,{ }{ c_4}'\,{ b_4}$& $ 1
  $ & $ 1
  $ \\[3mm]
$ { }{ c_3}'\,{ }{ c_3}\,{ b_4}$& $ -3\,{\sqrt{{{16864}\over
{79189}}}}
  $ & $ {{{\sqrt{{{621224}\over 5}}}}\over {253}}
  $ \\[3mm]
$ { }{ c_4}^{(3)}\,{ }{ c_4}\,{ b_4}$& $ {{1703}\over {3\,
{\sqrt{1335443296}}}}
  $ & $ {{47}\over {{\sqrt{3106120}}}}
  $ \\[1mm]
$ { }{ c_4}''\,{ }{ c_4}'\,{ b_4}$& $ {{1013}\over {3\,
{\sqrt{5341773184}}}}
  $ & $ {{40171}\over {13\,{\sqrt{12424480}}}}
  $ \\[3mm]
$ { }T\,{ }{ c_4}'\,{ }{ c_4}\,{ b_4}$& $ 0
  $ & $ {{-79\,{\sqrt{{{160}\over {77653}}}}}\over {13}}
  $ \\[3mm]
$ { }{ b_2}'\,{ }{ b_2}\,
    { }{ c_3}^{(3)}\,{ }{ c_3}\,{ c_4}$& $ {{5947 - 84994560\,\beta
    }\over {45\,{\sqrt{341873483776}}}}
  $ & $ {{-1967633671 + 11758527872\,\beta }\over
    {128271\,{\sqrt{795166720}}}}
  $ \\[1mm]
$ { }{ b_2}'\,{ }{ b_2}\,
    { }{ c_3}''\,{ }{ c_3}'\,{ c_4}$& $ {{-\left( 102617195 +
    15961978368\,\beta  \right) }\over
    {63\,{\sqrt{33493003332050944}}}}
  $ & $ {{144013865183 + 176377918080\,\beta }\over
    {897897\,{\sqrt{19879168000}}}}
  $ \\[1mm]
$ { }{ b_2}'\,{ }{ b_2}\,
    { }{ c_3}''\,{ }{ c_3}\,{ c_4}'$& $ {{421962613 +
    239429675520\,\beta }\over
    {315\,{\sqrt{33493003332050944}}}}
  $ & $ {{-\left( 57786879077 + 529133754240\,\beta  \right) }\over
    {897897\,{\sqrt{19879168000}}}}
  $ \\[1mm]
$ { }{ b_2}'\,{ }{ b_2}\,
    { }{ c_3}'\,{ }{ c_3}\,{ c_4}''$& $ {{-16580597 + 15961978368\,
    \beta }\over
    {63\,{\sqrt{33493003332050944}}}}
  $ & $ {{7500539549 - 13567532160\,\beta }\over
    {69069\,{\sqrt{19879168000}}}}
  $ \\[1mm]
$ { }{ b_2}''\,{ }{ b_2}'\,
    { }{ c_3}'\,{ }{ c_3}\,{ c_4}$& $ {{-497867}\over {9\,
    {\sqrt{523328177063296}}}}
  $ & $ {{164168647}\over {128271\,{\sqrt{12424480}}}}
  $ \\[3mm]
\end{tabular}
\end{center}
\newpage
\begin{center}
\begin{tabular}{l|l|l}
$ { }T\,{ }{ b_2}'\,
    { }{ b_2}\,{ }{ c_3}'\,{ }{ c_3}\,{ c_4}$& $
    {{{\sqrt{{{32384}\over {16160084519}}}}}\over 3}
  $ & $ {{2089\,{\sqrt{{{40}\over {77653}}}}}\over {3289}}
  $ \\[3mm]
$ { }{ W_3}\,{ }{ b_2}'\,
    { }{ b_2}\,{ }{ c_3}\,{ }{ c_4}'\,{ c_4}$& $ {{622159409 +
    747848245760\,\beta }\over {373924122880}}
  $ & $ {{-2487841011 + 58792639360\,\beta }\over {29396319680}}
  $ \\[3mm]
$ { }{ b_2}'\,{ }{ b_2}\,
    { }{ c_4}^{(4)}\,{ }{ c_4}'\,{ c_4}$& $ {{1942860007649 +
    833561113674240\,\beta }\over
    {567\,{2^{{{27}\over 2}}}\,{\sqrt{253}}\,{{164951}^{{3\over 2}}}}}
  $ & $ {{-\left( 85903480307899 + 363044548048000\,\beta  \right)
  }\over
    {3549\,{{49697920}^{{3\over 2}}}}}
  $ \\[1mm]
$ { }{ b_2}'\,{ }{ b_2}\,
    { }{ c_4}^{(3)}\,{ }{ c_4}''\,{ c_4}$& $ {{-51906446611 +
    25140115929600\,\beta }\over
    {81\,{2^{{{25}\over 2}}}\,{\sqrt{253}}\,{{164951}^{{3\over 2}}}}}
  $ & $ {{-84802144919311 + 10195878310450560\,\beta }\over
    {46137\,{2^{{{19}\over 2}}}\,{{388265}^{{3\over 2}}}}}
  $ \\[1mm]
$ { }{ b_2}''\,{ }{ b_2}'\,
    { }{ c_4}''\,{ }{ c_4}'\,{ c_4}$& $ {{-17078068387}\over
    {81\,{2^{{{19}\over 2}}}\,{\sqrt{253}}\,{{164951}^{{3\over 2}}}}}
  $ & $ {{9160909287863}\over
    {6591\,{2^{{{13}\over 2}}}\,{{388265}^{{3\over 2}}}}}
  $ \\[3mm]
$ { }T\,{ }{ b_2}'\,
    { }{ b_2}\,{ }{ c_4}''\,{ }{ c_4}'\,{ c_4}$& $ {{4143551}\over
    {27\,{\sqrt{506}}\,{{164951}^{{3\over 2}}}}}
  $ & $ {{-\left( 6276667282831 + 67811430237824\,\beta  \right)
  }\over
    {15379\,{2^{{{13}\over 2}}}\,{\sqrt{5}}\,{{77653}^{{3\over 2}}}}}
  $ \\[3mm]
$ { }{ b_2}\,{ }{ c_3}\,
     { }{ b_3}^{(3)}\,{ }{ c_4}'\,{ c_4}$& $ {{\left( 2291791 +
     8867765760\,\beta  \right) }\over {5320659456}}
  $ & $ {{100359111 + 3094349440\,\beta }\over {1856609664}}
  $ \\[1mm]
$ { }{ b_2}\,{ }{ c_3}'\,
     { }{ b_3}''\,{ }{ c_4}'\,{ c_4}$& $ {{5289}\over {659804}}
  $ & $ {{34701759}\over {52493428}}
  $ \\[1mm]
$ { }{ b_2}\,{ }{ c_3}\,
     { }{ b_3}''\,{ }{ c_4}''\,{ c_4}$& $ {{-\left(
31075 - 57211392\,\beta\right)
     }\over {14302848}}
  $ & $ {{-1979773899 + 58792639360\,\beta }\over {14698159840}}
  $ \\[1mm]
$ { }{ b_2}\,{ }{ c_3}''\,
     { }{ b_3}'\,{ }{ c_4}'\,{ c_4}$& $ {{-\left(
8701813 + 26603297280\,
     \beta \right)}\over {5320659456}}
  $ & $ {{5\,\left( 1141980071 - 11758527872\,\beta  \right)
  }\over
    {11758527872}}
  $ \\[1mm]
$ { }{ b_2}\,{ }{ c_3}'\,
     { }{ b_3}'\,{ }{ c_4}''\,{ c_4}$& $ {{\left( 43797 +
     23838080\,\beta  \right) }\over {2979760}}
  $ & $ {{-1146814353 + 29396319680\,\beta }\over {3674539960}}
  $ \\[1mm]
$ { }{ b_2}\,{ }{ c_3}\,
     { }{ b_3}'\,{ }{ c_4}^{(3)}\,{ c_4}$& $ {{-\left(135509501 +
     26603297280\,\beta \right)}\over {26603297280}}
  $ & $ {{2676877295 - 11758527872\,\beta }\over {11758527872}}
  $ \\[1mm]
$ { }{ b_2}\,{ }{ c_3}\,
     { }{ b_3}'\,{ }{ c_4}''\,{ c_4}'$& $ {{-\left(62263121 -
     79809891840\,\beta \right)}\over {13301648640}}
  $ & $ {{-1269768317 + 13567532160\,\beta }\over {2261255360}}
  $ \\[1mm]
$ { }{ b_2}\,{ }{ c_3}^{(3)}\,
     { }{ b_3}\,{ }{ c_4}'\,{ c_4}$& $ {{-\left(2060917 + 8867765760\,
     \beta \right)}\over {1330164864}}
  $ & $ {{177222537 - 4522510720\,\beta }\over {678376608}}
  $ \\[1mm]
$ { }{ b_2}\,{ }{ c_3}''\,
     { }{ b_3}\,{ }{ c_4}''\,{ c_4}$& $ {{-\left(
30543 - 591184384\,\beta\right)
     }\over {591184384}}
  $ & $ {{7914129141 + 58792639360\,\beta }\over {58792639360}}
  $ \\[1mm]
$ { }{ b_2}\,{ }{ c_3}'\,
     { }{ b_3}\,{ }{ c_4}^{(3)}\,{ c_4}$& $ {{689443}\over
     {158352960}}
  $ & $ -{{4322106}\over {13123357}}
  $ \\[1mm]
$ { }{ b_2}\,{ }{ c_3}'\,
     { }{ b_3}\,{ }{ c_4}''\,{ c_4}'$& $ {{\left( 7654343 +
     8867765760\,\beta  \right) }\over {1266823680}}
  $ & $ {{1780415853 + 58792639360\,\beta }\over {8398948480}}
  $ \\[1mm]
$ { }{ b_2}\,{ }{ c_3}\,
     { }{ b_3}\,{ }{ c_4}^{(4)}\,{ c_4}$& $ {{-\left(206198941 +
     168487549440\,\beta \right)}\over {53206594560}}
  $ & $ {{118119834753 - 1117060147840\,\beta }\over {352755836160}}
  $ \\[1mm]
$ { }{ b_2}\,{ }{ c_3}\,
     { }{ b_3}\,{ }{ c_4}^{(3)}\,{ c_4}'$& $ {{-\left(408011 -
     8867765760\,\beta\right) }\over {1662706080}}
  $ & $ {{-306567231 + 7349079920\,\beta }\over {1377952485}}
  $ \\[3mm]
$ { }T\,{ }{ b_2}'\,
     { }{ c_3}\,{ }{ b_3}\,{ }{ c_4}'\,{ c_4}$& $ {{\left( 6055439
     + 8867765760\,\beta  \right) }\over {2216941440}}
  $ & $ {{-3771447139 + 58792639360\,\beta }\over {14698159840}}
  $ \\[1mm]
$ { }T\,{ }{ b_2}\,
     { }{ c_3}\,{ }{ b_3}'\,{ }{ c_4}'\,{ c_4}$& $ {{\left(
     17862797 + 26603297280\,\beta  \right) }\over {13301648640}}
  $ & $ {{-5064153547 + 58792639360\,\beta }\over {29396319680}}
  $ \\[1mm]
$ { }T\,{ }{ b_2}\,
     { }{ c_3}'\,{ }{ b_3}\,{ }{ c_4}'\,{ c_4}$& $ {{\left(
     10285679 + 8867765760\,\beta  \right) }\over {4433882880}}
  $ & $ {{-16822056571 + 58792639360\,\beta }\over {29396319680}}
  $ \\[1mm]
$ { }T\,{ }{ b_2}\,
     { }{ c_3}\,{ }{ b_3}\,{ }{ c_4}''\,{ c_4}$& $ {{\left(
     1434213 + 2955921920\,\beta  \right) }\over {1477960960}}
  $ & $ {{673172789 + 58792639360\,\beta }\over {29396319680}}
  $ \\
\end{tabular}
\end{center}
\newpage
\begin{center}
\begin{tabular}{l|l|l}
$ { }{ b_2}\,{ }{ c_3}'\,
     { }{ c_3}\,{ }{ c_4}\,{ b_4}'$& $ {{12}\over {313}}
  $ & $ -{{17037}\over {16445}}
  $ \\[1mm]
$ { }{ b_2}\,{ }{ c_3}''\,
     { }{ c_3}\,{ }{ c_4}\,{ b_4}$& $ 0
  $ & $ 0
  $ \\[1mm]
$ { }{ b_2}\,{ }{ c_3}'\,
     { }{ c_3}\,{ }{ c_4}'\,{ b_4}$& $ {{16}\over {313}}
  $ & $ -{{22716}\over {16445}}
  $ \\[3mm]
$ { }{ b_2}\,{ }{ c_4}''\,
     { }{ c_4}'\,{ }{ c_4}\,{ b_4}'$& $ {{-\left(
15931409 + 19509084672\,
     \beta \right)}\over {1773553152}}
  $ & $ {{6371092053 - 129343806592\,\beta }\over {11758527872}}
  $ \\[1mm]
$ { }{ b_2}\,{ }{ c_4}^{(3)}\,
     { }{ c_4}'\,{ }{ c_4}\,{ b_4}$& $ {{-\left(
4061585 + 5320659456\,
     \beta \right)}\over {1330164864}}
  $ & $ {{1332601887 - 11758527872\,\beta }\over {2939631968}}
  $ \\[3mm]
$ { }{ c_3}'\,{ }{ c_3}\,
    { }{ b_3}'\,{ }{ b_3}\,{ c_4}$& $ {{-647}\over
    {{\sqrt{1335443296}}}}
  $ & $ {{-8559}\over {13\,{\sqrt{3106120}}}}
  $ \\[3mm]
$ { }{ b_3}'\,{ }{ b_3}\,
    { }{ c_4}''\,{ }{ c_4}'\,{ c_4}$& $ {{-25\,{{{{313}\over
    {527}}}^{{3\over 2}}}\,
      {\sqrt{{{23}\over {11}}}}}\over {9\,{2^{{{17}\over 2}}}}}
  $ & $ {{17457\,{5^{{7\over 2}}}}\over
    {{2^{{{11}\over 2}}}\,{{77653}^{{3\over 2}}}}}
  $ \\[3mm]
$ { }{ c_3}\,{ }{ b_3}\,
    { }{ c_4}'\,{ }{ c_4}\,{ b_4}'$& $ {{2789}\over
    {{\sqrt{1335443296}}}}
  $ & $ {{48249}\over {13\,{\sqrt{3106120}}}}
  $ \\[1mm]
$ { }{ c_3}'\,{ }{ b_3}\,
    { }{ c_4}'\,{ }{ c_4}\,{ b_4}$& $ {{1259}\over
    {{\sqrt{333860824}}}}
  $ & $ {{297\,{\sqrt{{{67}\over {11590}}}}}\over {13}}
  $ \\[1mm]
$ { }{ c_3}\,{ }{ b_3}\,
    { }{ c_4}''\,{ }{ c_4}\,{ b_4}$& $ 9\,{\sqrt{{{34}\over
    {2454859}}}}
  $ & $ {{567\,{\sqrt{{{40}\over {77653}}}}}\over {13}}
  $ \\[3mm]
$ { }{ b_2}''\,{ }{ b_2}'\,
    { }{ b_2}\,{ }{ c_3}'\,
      { }{ c_3}\,{ }{ c_4}'\,{ c_4}$& $ {{-\left( 994248158327 +
      1393241281850880\,\beta  \right) }\over
    {2106688508305920}}
  $ & $ {{3\,\left( -5374570365239 + 87024864780672\,\beta
  \right) }\over
    {27624141550720}}
  $ \\[3mm]
$ { }{ b_2}'\,{ }{ b_2}\,
     { }{ c_3}\,{ }{ b_3}'\,
       { }{ c_4}''\,{ }{ c_4}'\,{ c_4}$& $ {{-\left(160494387499 +
       176832117020160\,\beta \right)}\over
    {135\,{2^{{{21}\over 2}}}\,{\sqrt{253}}\,{{164951}^{{3\over 2}}}}}
  $ & $ {{-\left( 175846788774911 + 2804350104832640\,\beta
  \right) }\over
    {15379\,{{12424480}^{{3\over 2}}}}}
  $ \\[1mm]
$ { }{ b_2}'\,{ }{ b_2}\,
     { }{ c_3}'\,{ }{ b_3}\,
       { }{ c_4}''\,{ }{ c_4}'\,{ c_4}$& $ {{-\left(423757135087 +
       110962352954880\,\beta \right)}\over
    {315\,{2^{{{23}\over 2}}}\,{\sqrt{253}}\,{{164951}^{{3\over 2}}}}}
  $ & $ {{-3\,\left( 71812352639835 + 69927965254784\,\beta
  \right) }\over
    {15379\,{2^{{{17}\over 2}}}\,{\sqrt{5}}\,{{77653}^{{3\over 2}}}}}
  $ \\[1mm]
$ { }{ b_2}'\,{ }{ b_2}\,
     { }{ c_3}\,{ }{ b_3}\,
       { }{ c_4}^{(3)}\,{ }{ c_4}'\,{ c_4}$& $ {{-\left(
38872420967 +
       62603470343680\,\beta\right) }\over
    {105\,{2^{{{21}\over 2}}}\,{\sqrt{253}}\,{{164951}^{{3\over 2}}}}}
  $ & $ {{22901871803033 - 306838784819840\,\beta }\over
    {15379\,{2^{{{11}\over 2}}}\,{{388265}^{{3\over 2}}}}}
  $ \\[3mm]
$ { }{ b_2}'\,{ }{ b_2}\,
    { }{ c_3}'\,{ }{ c_3}\,
      { }{ c_4}'\,{ }{ c_4}\,{ b_4}$& $ {{-\left( 13632557 +
      26603297280\,\beta  \right) }\over
    {105\,{\sqrt{2093312708253184}}}}
  $ & $ {{-9669258947 + 58792639360\,\beta }\over
    {299299\,{\sqrt{1242448000}}}}
  $ \\[3mm]
$ { }{ b_2}\,{ }{ c_3}'\,
    { }{ c_3}\,{ }{ b_3}'\,
      { }{ b_3}\,{ }{ c_4}'\,{ c_4}$& $ {{4424047 + 8867765760\,
      \beta }\over {886776576}}
  $ & $ {{-855872103 + 4522510720\,\beta }\over {452251072}}
  $ \\[3mm]
$ { }{ b_2},{ }{ c_3}\,
     { }{ b_3}\,{ }{ c_4}''\,
       { }{ c_4}'\,{ }{ c_4}\,{ b_4}$& $ {{-\left(
7808239 + 8867765760\,
       \beta\right) }\over {1108470720}}
  $ & $ {{-\left( 1624519269 + 58792639360\,\beta  \right) }\over
    {7349079920}}
  $ \\[3mm]
\end{tabular}
\end{center}

\end{document}